\documentclass[11pt,a4paper]{article}
\pdfoutput=1

\usepackage{jcappub} 

\usepackage[T1]{fontenc}
\usepackage{amssymb,amsmath}
\usepackage{bm}

\usepackage{footnote}

\usepackage{bm}

\usepackage{hyperref}

\usepackage{amssymb,amsmath}

\newcommand{\ud}{\,\mathrm{d}}
\newcommand{\uD}{\,\text{D}}

\usepackage{xcolor}

\newcommand{\beq}{\begin{equation}}
\newcommand{\beqn}{\begin{eqnarray}}
\newcommand{\eeq}{\end{equation}}
\newcommand{\eeqn}{\end{eqnarray}}

\usepackage{physics}

\usepackage{multirow}

\newcommand{\sgn}{\mathop{\mathrm{sgn}}}

\usepackage{verbatim}

\begin{document}
 
\title{\Huge  Attractors, Bifurcations and Curvature  in Multi-field Inflation}
\author[a]{Perseas Christodoulidis,}
\author[a]{Diederik Roest,}
\author[b,c]{Evangelos I. Sfakianakis}
\affiliation[a]{Van Swinderen Institute for Particle Physics and Gravity, 
University of Groningen, Nijenborgh 4, 9747 AG Groningen, The Netherlands}
\affiliation[b]{Nikhef, Science Park 105, 1098 XG Amsterdam, The Netherlands}
\affiliation[c]{Lorentz Institute for Theoretical Physics, Leiden University, 2333 CA Leiden, The Netherlands}
\emailAdd{p.christodoulidis@rug.nl}
\emailAdd{d.roest@rug.nl}
\emailAdd{e.sfakianakis@nikhef.nl}

\abstract{Recent years have seen the introduction of various multi-field inflationary scenarios in which the  curvature and geodesics of the scalar manifold play a crucial role. We outline a simple description that unifies these different proposals and discuss their stability criteria. We demonstrate how the underlying dynamics is governed by an effective potential, whose critical points and bifurcations determine the late-time behaviour of the system, thus unifying hyperinflation, angular, orbital and side-tracked inflation. Interestingly, we show that hyperinflation is a special case of side-tracked inflation, relying on the enhanced isometries of the hyperbolic manifold. We provide the explicit coordinate transformation that maps the two models into each other. Finally, we relax the assumption of a field-space isometry along the inflationary direction that has been considered a prerequisite in the literature so far. We explicitly construct inflationary solutions that do not proceed along a field-space isometry or geodesic and use them to discuss stability criteria.}

\maketitle

\tableofcontents

\section{Introduction} 

Inflation, the hypothesis of rapid accelerated expansion in the primordial Universe, provides an elegant solution to the flatness and horizon problems \cite{Guth, Linde}, and seeds the primordial Universe with quantum fluctuations whose predictions 
are in excellent agreement with the latest CMB observations \cite{Planck}. 
Inflation requires $\epsilon<1$, where $\epsilon =- {\ud}(\log H)/\ud N $  
is the first Hubble flow (or slow roll) parameter and $N$ is the number of e-folds. Quasi-De Sitter expansion requires $\epsilon\ll1$, and hence requires a small deviation from the scale invariant De Sitter space-time for a prolonged period of time. 
Prolonged inflation imposes restrictions on the evolution of $\epsilon$,  as a function of e-folds
$ \epsilon' \ll 1$, where $'\equiv d/dN$, independently of the requirement that $\epsilon$ should be small. 
Note that this condition is weaker (and more general) than the one usually used, namely $\eta \equiv \epsilon' / \epsilon$ being small, since $\epsilon<1$ during inflation.

We consider a model consisting of multiple scalar fields $\Phi^I$, $I\in [1,\mathcal{N}]$, and non-canonical kinetic terms 
\begin{equation}
\mathcal{L} = \sqrt{-g} \left( {1 \over 2} R - {1\over 2}{\cal{G}}_{IJ}(\Phi^K) \partial_\mu\Phi^I \partial^\mu\Phi^J  -V \right) \, ,
\end{equation}
(we use units of $M_{\rm Pl} =1$) where $\mu,\nu$ are spacetime indices and the matrix ${\cal{G}}_{IJ}$ can be interpreted as a field-space metric\footnote{Our subsequent analysis can be applied in other types of multi-scalar systems, such as the holographic RG flow of domain walls, under an appropriate mapping (see e.g. Refs.~\cite{Skenderis:2007sm,McFadden:2009fg}).}. The first slow-roll parameter can be written as $\epsilon =  v_I v^I/2$,
where $v^I\equiv \ud \Phi^I / \ud N $ is  velocity of the scalar fields with respect to the $e$-folding number, $N$. The latter is related to cosmic time via the Hubble parameter $\ud N = H \ud  t $ defined as
\begin{equation} \label{Hubble}
3 H^2 =  \tfrac12 \mathcal{G}_{IJ} \partial_t{\Phi}^I \partial_t{\Phi}^J + V \, .
\end{equation}
The variation of $\epsilon$ reads
 \begin{equation}
\epsilon'= v_I a^I \,, \quad a^I \equiv \uD_N v^I = \frac{\ud v^I}{\ud N} + \Gamma^I_{JK} v^J v^K \,,
  \label{eq:eta}
 \end{equation}
where $a^I$ is the covariant (or generalised) acceleration and $\Gamma^I_{JK}$ the Christoffel symbols associated with ${\cal G}_{IJ}$.
 We  express the scalar field
  equations of motion as
    \begin{equation} \label{fieldeqs}
   a^I = - (3 - \epsilon)  v^I -  {V^{,I} \over H^2}  \,,
  \end{equation}
where the RHS consists of the Hubble friction and the potential gradient terms $V^{,I} = {\cal G}^{IJ}\partial V/\partial \Phi^J$. 
   
Prolonged inflation requires $\epsilon$ to be approximately constant, translating into the (approximate) vanishing of $\epsilon'$, the inner product between the velocity and covariant acceleration of the  fields \eqref{eq:eta}. For a single field, this implies that the acceleration must be very small and that it is necessary for  prolonged single-field inflation  to impose the slow-roll condition $\ddot \phi \approx 0$, given by the separate vanishing of the two sides of the scalar field equation \eqref{fieldeqs}. 
Fast-roll inflation can be achieved by including higher-order derivative terms as in e.g.~DBI inflation \cite{DBI}. 

A simple generalization in multi-field inflation is the slow-roll slow-turn condition, restricted to potential gradient flow (see e.g. Refs.~\cite{GrootNibbelink:2000vx, GrootNibbelink:2001qt, Peterson:2011yt, Lyth:2009zz, Yang:2012bs}), where $\epsilon$  and the covariant acceleration vector $a^I$ are both small.
However, $\epsilon'$ can be vanishing while some of the components $a^I$ are large, as long as the acceleration is perpendicular to  $v^I$.
This requires an interplay between gradient terms $\propto V^{,I}$ and (generalised) centrifugal forces  $\propto \Gamma^I_{JK} v^J v^K$. We will explicitly demonstrate this by constructing an {\it effective potential}, that can be linked to the Hubble parameter, and show that it can describe dynamics shared by all recent models that exhibit a novel inflationary attractor 
\cite{Achucarro:2015rfa, Christodoulidis:2018qdw, Linde:2018hmx,  Garcia-Saenz:2018ifx, Achucarro:2019pux}. While hyperinflation \cite{Brown:2017osf,Mizuno:2017idt} might appear to be of a different nature, it is also  captured by our effective potential formalism. 
We show that  hyperinflation is a special case of sidetracked inflation, going
beyond recent investigations that have pointed out similarities between them \cite{Bjorkmo:2019aev,Fumagalli:2019noh,Bjorkmo:2019fls} in the context of geometrical destabilization  \cite{Renaux-Petel:2015mga,Renaux-Petel:2017dia,Cicoli:2018ccr,Grocholski:2019mot}. 
 
The paper is organized as follows: in Sec~\ref{sec:be} we demonstrate how all recent novel attractor solutions can collectively be described as the late-time evolution in a special coordinate system, where all fields but one are non-dynamical. This coordinate choice will allow us to express the attractor solution in a coordinate invariant form. In Sec~\ref{sec:stab_bif} we derive the stability conditions for the background solution and then focus on bifurcations between different solutions. Analyzing the bifurcation structure of hyperinflation we show that it belongs to the sidetracked family of models. In Sec~\ref{sec:iso_pert} we contrast our stability criteria with previous conditions found in the literature and briefly discuss quantum fluctuations of the novel attractor models. We offer our conclusions in Sec~\ref{sec:conclusion}.

\section{Background evolution}\label{sec:be}

\subsection{Late-time dynamics}
The `typical' evolution of many multi-field inflationary models 
consists of an early period of multi-field behavior 
and a late period of single-clock inflation. 
Depending on the duration of each phase, the relevant part of the evolution (last $50-60$ $e$-folds) is described by an ${\cal N}$-dimensional hypersurface, where ${\cal N}$ is the number of evolving degrees of freedom, or a single trajectory if all other ``orthogonal'' fields have relaxed to their minimum. In this letter we mainly focus  on the last phase and so we decompose  the scalar fields $ \Phi^I = (\phi, \chi^i)$, where $\phi$ is defined as the (light) inflationary direction and $\chi^i$ are the orthogonal fields that are (approximately) constant during inflation. This split is manifested in an appropriate coordinate system where $(\chi^i) ' \approx 0$ will hold as an (approximate) solution.  
 
Specializing to $\mathcal{N}=2$ (though this argument also holds for an arbitrary number of fields) a given solution $\{(\psi^{1}_{\rm sol})',(\psi^{2}_{\rm sol})'\}$, where the velocity components can be non-zero in general, can be mapped to $\{(\phi^{1}_{\rm sol})',0\}$ under the coordinate transformation which as usual transforms the components as
\begin{align}
	(\phi^I_{\rm sol})' &= {\partial \phi^I \over \partial \psi^K} (\psi^{K}_{\rm sol})' \, .
\end{align}
Since the existence of an attractor is assumed, velocities are given as functions of the fields, and the partial differential equation for the unknown function $\phi^2$
\begin{equation}
{\partial \phi^2 \over \partial \psi^1} (\psi^{1}_{\rm sol})' + {\partial \phi^2 \over \partial \psi^2} (\psi^{2}_{\rm sol})' =0 \, ,
\end{equation}
has the form of the advection equation with variable coefficients. This can always be solved (for instance with the method of characteristics) and this proves the existence of our coordinate construction.

This coordinate choice leads to\footnote{Note that the present construction differs from the adiabatic/entropic decomposition \cite{Kaiser:2012ak,  Peterson:2011yt, Renaux-Petel:2015mga, Achucarro:2012sm, Gong:2011uw} since the latter does not introduce a new coordinate system. Instead, the adiabatic direction is related to our inflationary direction as $\dot \sigma^2 = {\cal G}_{\phi\phi} \dot\phi^2$.}
 \begin{equation} \label{velocities}
  v^I \approx (v, 0, \ldots) \,,~ a^I \approx \left(\frac{\ud v}{\ud N} + \Gamma^\phi_{\phi \phi} v^2, \Gamma^i_{\phi \phi} v^2 \right) \,, 
 \end{equation}
evaluated on that particular inflationary solution.
While field-space manifolds with isometries provide natural choices for this parametrization \cite{Achucarro:2019pux},  we   show that the presence of an isometry is not necessary. Also, the isometry structure of hyperbolic space allows for different equivalent parametrizations.

We observe a particularly striking separation of the consequences of prolonged inflation  ($\epsilon'\ll1$). Along the inflationary direction $v \uD_N v \ll 1$, which through the equation of motion \eqref{fieldeqs} yields $ v \approx -V^{,\phi}/V$. 
This implies that the inflationary direction is subject to the usual slow-roll condition, where Hubble friction is balanced by the potential gradient. For generic potentials consistency of this solution requires the smallness of first and second slow-roll parameters in the inflationary direction. In our coordinate system the two conditions read (see App.~\ref{app1}): 
\begin{equation} \label{eq:sr}
\frac{1}{2}\mathcal{G}^{\phi\phi} \left( {V_{,\phi} \over V}\right)^2 \ll 1 \, ,\qquad 
\mathcal{G}^{\phi\phi} { V_{,\phi\phi} \over V }  \ll  1 \, .
\end{equation}

The situation is radically different for the orthogonal field directions. By adapting our coordinates, we have defined these as stationary that can  have a non-vanishing covariant acceleration  only when deviating away from a geodesic. This  introduces a (generalized) centrifugal force that is balanced by a potential gradient: for the stationary directions Eqs.~\eqref{fieldeqs} become
 \begin{equation} \label{stationary}
  V_{\rm eff}^{,i} \equiv  V^{,i} + \Gamma^i_{\phi \phi} v^2 H^2 \approx 0 \,.
 \end{equation}
We call this the {\it effective gradient} along the $i$'th direction in field space. Note that contrary to the inflationary direction, consistency of these conditions imposes no restrictions on $V^{,i}$ (apart from having different signs with respect to $\Gamma^i_{\phi \phi}$). 
This decouples the potential gradient from the inflationary trajectory, providing the means to evade the refined de Sitter conjecture of Ref.~\cite{Ooguri:2018wrx}.

Eqs.~\eqref{stationary} should be seen as algebraic relations for the stationary fields $\chi^i$, in terms of the inflaton field $\phi$ and its velocity $v$.  The stationary fields will adapt their values to balance the centrifugal and potential forces acting on them, as in the gelaton model \cite{Tolley:2009fg}. Therefore, at a given moment during inflation, i.e.~for a particular value of $\phi$, one can view Eq.~\eqref{stationary} as the gradient of  an effective potential, whose extrema fix the values of these fields, akin to moduli stabilisation in string theory. When both terms in the right hand side of Eq.~\eqref{stationary} vanish separately, one has slow-roll slow-turn conditions for potential gradient flow, which  is by no means necessary in the multi-field case.  In general, negative curvature tends to induce non-geodesic motion.

There is an attractive interpretation of the above condition when formulated in phase space. The effective potential \eqref{stationary} coincides with the total energy (and the Hubble parameter) as a function of the orthogonal field values $\chi^i$, for a given value of the inflaton $\phi$ and its conjugate momentum $\pi_\phi = {\cal G}_{\phi \phi} \dot \phi$. 
 In other words, the space-time metric and the inflaton field are assumed as a fixed time-dependent background, and the orthogonal fields are subject to the energy extremization condition 
 \begin{equation}
   \partial_i \left( \frac{1}{2}{\cal G}^{\phi \phi} (\phi, \chi^i) \pi_\phi^2 + V(\phi,\chi^i) \right) = 0 \,.
 \end{equation}
The orthogonal field dependence of the first term comes in via ${\cal G}^{\phi \phi}$ which, for negative curvature manifolds, decreases as one moves away from the geodesic solution with $\partial_i {\cal G}_{\phi \phi} = 0$. This allows for a competition between an increase in potential and a decrease in kinetic energies, providing an intuitive interpretation of geometric destabilization \cite{Renaux-Petel:2015mga,Renaux-Petel:2017dia,Cicoli:2018ccr,Grocholski:2019mot} as a simple competition of  energy contributions.

\subsection{Conditions and examples}

We  now derive a coordinate independent expression for the attractor solution in the case of two fields. For any two-dimensional metric, off-diagonal componentis can be set to zero with an appropriate redefinition of either $\chi$ or $\phi$.  By redefining $\phi$, the solution $\chi' \approx 0$ carries over in the new system and so without loss of generality we can assume the following diagonal metric
\begin{equation} 
\ud s^2 =  \mathcal{G}_{\phi \phi}(\phi,\chi) \ud \phi^2 + \mathcal{G}_{\chi \chi} (\phi,\chi) \ud \chi^2 \,. 
\end{equation} 
The solution for the slow-turn limit is already in covariant form, $\epsilon\approx \epsilon_V$, thus we will focus on the case when $\omega/H \equiv \Omega $ is non-negligible. Simple and manifestly invariant expressions, including only covariant derivatives of the potential, are the norm of the potential gradient, the trace of the Hessian and the projection of the Hessian along the potential gradient. As we will show shortly the previous three quantities suffice to derive a coordinate independent expression for the attractor solution. Our calclulations will drastically simplify using the coordinate system we defined earlier and we will further assume the slow-roll conditions for prolonged inflation $\epsilon, |\eta| \ll 1$.

\begin{enumerate}

\item The norm of the potential gradient is
\begin{equation} \label{eq:eps_turn}
	{\mathcal{G}^{IJ} V_{,I}V_{,J} \over V^2}\equiv 2 \epsilon_V \approx 2 \epsilon  + {2 \epsilon H^2 \omega^2 \over V^2} \Rightarrow \epsilon_V \approx \epsilon \left(1 + {\Omega^2 \over 9 }\right)\, ,
\end{equation}  
where the latter was first derived in Ref.~\cite{Hetz:2016ics}. As a side-note, the Swampland conjectures constrain the norm of the potential gradient and thus the above equation shows how one can have slow roll inflation with $\epsilon \ll 1$ on a steep potential with $\epsilon_v \ge{\cal O}(1)$.

\item The trace of the Hessian is
\begin{equation} \label{eq:dv}
	C_2 \equiv {\mathcal{G}^{IJ}V_{;IJ} \over V } =  {V_{;\sigma \sigma} \over V } + {V_{;ss} \over V }  \, .
\end{equation}
where we used the relation ${\cal G}^{IJ} = \hat \sigma^I \hat \sigma^J + \hat s^I \hat s^J$, which holds for two fields \cite{Kaiser:2012ak}, along with the definitions 
$V_{;\sigma \sigma}  = \hat \sigma^I \hat \sigma^J V_{;IJ}$ and $V_{;s s}  = \hat s^I \hat s^J V_{;IJ}$. Note that neither   ${V_{;ss} / V }$ nor ${V_{;\sigma \sigma} / V }$ need be small in order to get successful slow-roll inflation. 

We can trade $V_{;\sigma\sigma}$ for the turn rate using the definition of the slow-roll parameter in the adiabatic direction $\eta_{\sigma \sigma} = {V_{;\sigma\sigma} \over V } - {1 \over 3} \Omega^2$ (see App.~\ref{app1}), and neglecting $\eta_{\sigma \sigma} $ \begin{equation}  
	C_2 \equiv {\mathcal{G}^{IJ}V_{;IJ} \over V } \approx  {1 \over 3} \Omega^2 + {V_{;ss} \over V }  \, .
\end{equation}
However, for certain models (e.g. the multi-field alpha attractors) $V_{;ss}$ is negative and the two projections of the potential almost cancel each other. Hence, the previous substitution is invalid as their difference is of the same order as $\eta_{\sigma\sigma}$. To include these cases as well we will use $V_{;\sigma\sigma}$ instead of the turn rate and rewrite Eq.~\eqref{eq:eps_turn} as
\begin{equation}
\epsilon_V \approx \epsilon \left( 1 + {1 \over 3}{V_{;\sigma\sigma} \over V }\right) \, ,
\end{equation}
assuming $\Omega^2\gg |\eta_{\sigma\sigma}|$.

\item The third curvature invariant is
\begin{align} \label{eq:ddv}
	C_3 & \equiv {V^{,I} V^{,J} V_{;IJ} \over V^3} = {2\epsilon \over 9}  \Omega^2 {V_{;ss} \over V }  + 2\epsilon {V_{;\sigma \sigma} \over V }+ 2 {V^{,\chi} V^{,\phi} V_{;\chi\phi} \over V^3} 
\approx 2 \epsilon {V_{;\sigma \sigma} \over V } \left( {1 \over 3} {V_{;ss} \over V }  -1\right)
\end{align}
where we related the time derivative of the turn rate with the mixed derivatives through
\begin{equation}
	{1\over 9} \epsilon (\Omega^2) ' = - {V^{,\chi} V^{,\phi} V_{,\chi\phi} \over  V^3} +  {V^{,\chi} V^{,\phi} \Gamma^{\chi}_{\phi\chi }V_{,\chi} \over V^3}  + {1\over 9}  \epsilon \Omega^2 \left(   4 \epsilon - \eta \right)  \, ,
\end{equation} 
and is thus suppressed.
\end{enumerate}

\noindent Eliminating $V_{;ss}/V$ and $V_{;\sigma\sigma}/V$ in the system of Eqs.~\eqref{eq:dv}-\eqref{eq:ddv} yields a quadratic equation for $\epsilon$
\begin{equation}\label{eq:qeps}
2C_2 \epsilon^2 + (C_3 - 6 \epsilon_V -2 C_2 \epsilon_V) \epsilon +6 \epsilon_V^2 = 0\, ,
\end{equation}
 which in general has two solutions.
We can further simplify the solution by noticing that the relations $\epsilon < \epsilon_V$ and $\epsilon \ll 1$, with no such restrictions on the other curvature invariants, allow us to neglect the quadratic term in Eq.~\eqref{eq:qeps} and obtain instead
\begin{equation} \label{eq:epsapprox}
\epsilon \approx {6 \epsilon_V^2 \over  6 \epsilon_V + 2 C_2 \epsilon_V - C_3} \, .
\end{equation}
It is worth mentioning the two sets of assumptions we made to derive the previous formula. Firstly, we assumed that the magnitude of the tangential and orthogonal directions of the potential  ($V_{;\sigma \sigma}/V$, or equivalently $\Omega^2$, and $V_{;ss}/V$) are free parameters, but non-negligible compared to $\epsilon$ and $\eta$; otherwise one recovers the traditional slow-roll slow-turn approximation. Secondly, we neglected logarithmic prime derivatives of various quantities (e.g. $\epsilon$ and $\Omega$), as it allows us to neglect second order time derivatives and subsequently make analytical approximations possible. {The latter was assumed in the early works, such as Ref.~\cite{Achucarro:2010da}, as a definition of slow-roll, as well as in the derivation of the rapid-turn solution in Ref.~\cite{Bjorkmo:2019fls}. A discussion of the validity of this assumption was presented in Ref.~\cite{Christodoulidis:2019jsx}, where various two-field models have been shown to be approximated by scaling solutions with adiabatically changing parameters, for which logarithmic time derivatives are identical zero.} Therefore, Eq.~\eqref{eq:epsapprox} may hold even at the slow-turn limit and describes every possible late-time evolution with the previous assumptions satisfied. Note though, that the slow-turn limit might be different from the gradient-flow approximation, where the covariant acceleration becomes subdominant; here instead motion can be non-geodesic and yet the turning rate might be small (we will illustrate this point using the angular inflation model). 

Now we can show that the novel slow-roll behaviour found in recent models
\cite{Achucarro:2015rfa, Christodoulidis:2018qdw, Linde:2018hmx, Garcia-Saenz:2018ifx, Achucarro:2019pux,Brown:2017osf,Mizuno:2017idt} are all captured by the previous solution. Our first example is hyperinflation, formulated on the Poincar\'{e} disc with a spherically symmetric potential  \cite{Brown:2017osf,Mizuno:2017idt}
\begin{equation}\label{disc}
	\ud s^2 =  L^2 \sinh^2 \left({\rho \over L}\right) \ud\theta^2 + \ud \rho^2  \,,  \qquad V=V(\rho) \, .
\end{equation}
The solution for $\rho \gg L$ and $V_{,\rho} \gg 3L V$ is
\begin{equation}
\epsilon \approx {3 \over 2} L {V_{,\rho} \over V} \, ,
\end{equation}
recovering  the usual hyperinflation result.
This was first derived in Ref.~\cite{Brown:2017osf} and subsequently extended in Refs.~\cite{Mizuno:2017idt, Bjorkmo:2019aev, Christodoulidis:2019jsx}. In the opposite limit of shallow potentials and / or mildly curved manifolds we get simple gradient flow evolution wilh $\epsilon=\epsilon_V$ (the precise stability criteria are explained in Sec.~\ref{sec:stab}).

The second example is a two-field generalization of $\alpha$-attractor models \cite{alpha-attractors},  
where the scalar potential takes a finite value at the boundary of  the Poincar\'{e} disc\footnote{The Poincar\'e disk possesses unit radius and constant negative curvature. For $\alpha$-attractors models the curvature reads $R=-8/\alpha$, where the parameter $\alpha$ is defined through the field-space metric given in Eq.~\eqref{eq:ang}. } and exhibits angular dependence.  The field metric and potential are
\begin{equation} \label{eq:ang}
\ud s^2 = {6 \alpha \over (1 - r^2)^2}\left( \ud r^2 + r^2 \ud \theta^2 \right) \, , \qquad  V = 3 \alpha r^2 \left( m_1^2 \cos^2 \theta + m_2^2 \sin^2 \theta \right) \, .
\end{equation}
Such models  proceed for a prolonged number of e-folds along a slow-roll, slow-turn trajectory, 
giving rise to the universal predictions of $\alpha$-attractors for intermediate field-space curvature \cite{Achucarro:2017ing}. 
For large negative curvature (small values of $\alpha$), trajectories collapse to a particular late-time attractor, where motion proceeds predominately along the angular direction. For values of the mass ratio $R_{\rm m} \equiv (m_2/m_1)^2 \gtrsim O(10)$ the duration of the angular phase becomes inversely proportional to the $\alpha$ parameter \cite{Christodoulidis:2018qdw}. Note that there are effectively 3 quantities in this model: the distance from the boundary of the Poincar\'{e} disc, parameterized by $1 - r^2$, the field-space curvature controlled by $\alpha$ and the mass ratio $R_m$, or equivalently the potential steepness along the angular direction $V_{,\theta}/V$. Assuming moderate mass ratio $R_{\rm m}  \lesssim O(10)$, i.e.~$V_{,\theta} \ll V$ and $V_{,\theta \theta} \ll V$, then expanding Eq.~\eqref{eq:epsapprox} around small angular gradient we obtain
\begin{equation} \label{eq:epsang1}
{\epsilon}\simeq \frac{6 \left(1-r^2\right)^2}{18 \alpha +4 \left(1-r^2\right)} \, .
\end{equation}
To further simplify the previous we need to distinguish between two cases, depending on the relative size of $\alpha$ to $1 - r^2$. Hence, we obtain
\begin{align} 
\epsilon &\approx   {\left(1-r^2\right)^2 \over 3 \alpha} = \epsilon_V ~~(1-r^2\ll  \alpha) \, , \\ \label{eq:ang_eps_lt}
\epsilon &\approx {3  \over 2} \left(1- r^2\right) ~~ (1-r^2\gg  \alpha) \, .
\end{align}
The latter equations have a simple interpretation: for a given $\alpha$ two solutions are possible depending on the distance relative to the boundary of the Poincar\'{e} disc. If fields start very close to the boundary then they will first follow an almost radial evolution with a small turn-rate, but as their distance from the origin decreases there will be a transition towards a different solution, the novel angular attractor, in accordance with the findings of Refs.~\cite{Achucarro:2017ing, Christodoulidis:2018qdw}. Turning to the large mass ratio, i.e.~$V_{,\theta} \gg V$ only the slow-turn solution is possible at $1-r^2 \ll \alpha$  with 
\begin{equation} \label{eq:ang_eps_st}
\epsilon = {\left(1-r^2\right)^2 \over 12 \alpha} \left({V_{,\theta} \over V} \right)^2 \approx \epsilon_V \, .
\end{equation}

Despite being a slow-turn solution, motion does not proceed along the potential flow, i.e.~$(\phi^I)' \sim V^{,I}$ which results into boomerang-like curves, but along the angular direction instead \footnote{This can be deduced as follows:  the requirement of one frozen field and one field in slow roll combined with the solution $\epsilon \approx \epsilon_{\theta}$  (where the latter denotes the $\theta$ part in the definition of $\epsilon_V$), gives $\theta' \approx \theta'_{\rm SR}$ and $r'\approx 0$. Thus, this particular slow-turn solution belongs to the angular inflation regime.}. 
Since we found two seemingly different solutions, both proceeding predominately along the angular direction, we can try to unify their description. Equating Eqs.~\eqref{eq:ang_eps_st} and \eqref{eq:ang_eps_lt} provides the parametric relation between $r$ and $\theta$ during angular inflation 
\begin{equation}
1-r^2 \approx18 \alpha \left({V_{,\theta} \over V} \right)^{-2}  
= {9 \alpha \over 2} {\left( \cot \theta +R_{\rm m} \tan \theta \right)^2 \over ({R_{\rm m} -1})^2} \, ,
\end{equation}
where the last part of the equation refers to  the particular potential choice given in Eq.~\eqref{eq:ang}.
Of course, following either Eq.~\eqref{eq:epsapprox} or the analysis of Ref.~\cite{Christodoulidis:2018qdw}, one can study angular inflation solutions in potentials with arbitrary angular and radial dependence.


\section{Stability and bifurcations} \label{sec:stab_bif}
\subsection{Stability conditions}  \label{sec:stab}

The stability conditions for a general background solution 
are determined by the eigenvalues of the full stability matrix spanned by the fields and their velocities. 
In the cases of interest in this paper, with $\epsilon' \ll 1$ and $\chi^i \simeq {\rm constant}$, the stability criteria are given by the expansion of the effective potential 
at quadratic order, i.e.~$\partial_{i} V^{,j}_{\rm eff}$, and  algebraic restrictions on $\epsilon$ (see Ref.~\cite{Christodoulidis:2019jsx} and App.~\ref{app} for more details). Since we substitute an approximate solution the conditions listed below will be accurate to first order in the slow-roll parameters.

For clarity we  consider the two-field case, where any metric can be diagonalised. Moreover, we  restrict ourselves to the following form for the field metric
\begin{equation} 
\ud s^2 =  \mathcal{G}_{\phi \phi}(\chi) \ud \phi^2 + \mathcal{G}_{\chi \chi} (\phi) \ud \chi^2 \,. \label{general-metric}
\end{equation} 
The expression for the Ricci scalar of this manifold splits in two parts, $R = R^{(\phi)} + R^{(\chi)}$, parametrizing the derivative dependence on the two fields (there are no mixed derivatives $\partial_\phi \partial_\chi$). Motion along the $\phi$ direction is stable, as long as 
\begin{equation}\label{eq:mvel}
  3 -\epsilon  + \left( \log \sqrt{{\cal G}}\right)' > 0\, ,
\end{equation}
where ${\cal G} \equiv {\rm det} \left ({\cal G}_{IJ}\right )$. 
We will provide the physical interpretation later using  hyperinflation \cite{Brown:2017osf} as an example. In addition, the effective mass (defined as the linearization of $V^{,\chi}_{\rm eff}$) reads
\begin{equation}
 M_{\rm eff}^2 =  V_{,\chi}{}^\chi  + \epsilon H^2 R^{(\chi)} + 3 \frac{V_{,\chi} V^{,\chi}}{2 \epsilon H^2}  \, .
\label{eq:ms}
\end{equation}
Since multi-field trajectories can deviate from the gradient flow, one can define the turn rate $\omega$ as the deviation of $\epsilon$ from $\epsilon_V \equiv \frac12(\log V)_{,I}(\log V)^{,I}$. For our choice of coordinates $\omega^2 = {V_{,\chi} V^{,\chi}}/({2 \epsilon H^2} )$.

An example with a single stable attractor is provided by two-field $\alpha$-attractor models modeled by Eq.~\eqref{eq:ang} and explored in detail in Ref.~\cite{Christodoulidis:2018qdw}. It is straightforward to check that the effective gradient $V_{\rm eff}^{,\rho}$ stabilizes the radius near the boundary of the Poincar\'e disc, leading to a late-time 
attractor with non-vanishing turn rate, proceeding along a non-geodesic direction in field space. 
 
Turning to a second example, it was recently pointed out \cite{Achucarro:2016fby, Achucarro:2019pux} that neutral stability can be achieved using the Hamilton-Jacobi formalism, where the scalar potential is given in terms of the Hubble parameter 
by 
\begin{equation}
V= 3 H^2 - 2 H_{,I} H^{,I} \, .
\end{equation} 
This formalism has an exact first-order solution for the scalar velocities  \cite{Salopek:1990jq} \footnote{This can be seen as the cosmological analogue of the first-order equation that governs AdS critical points and BPS domain walls \cite{Townsend, Skenderis}.} 
\begin{equation}
v^I = -2 H^{,I} / H \, . 
\end{equation}  
Upon adapting coordinates such that $H=H(\phi)$, one has a natural distinction between the inflationary and the stationary directions. Such trajectories may be (strongly) turning, however, as the Hubble gradient may differ from the potential gradient. The latter will be non-vanishing if the metric along the inflationary direction $\mathcal{G}_{\phi \phi}$ depends on the stationary directions, resulting in
 \begin{equation}
 \label{eq:HJ}
   V^{,i} = -2 \partial^i {\cal G}_{\phi \phi} (H^{,\phi})^2 \,,
 \end{equation}
which is equivalent to the vanishing of the effective gradient of Eq.~\eqref{stationary}. The latter is therefore identically satisfied, leading to neutrally stable stationary points and hence flat directions in the effective potential and  Hubble parameter, which are directly related to the choice $H=H(\phi)$. This implies that the field space is spanned by adjacent trajectories. One thus has a convergence of the $2n$-dimensional phase space of initial conditions to the $n$-dimensional hypersurface that fixes the fields' velocity but not their positions.

\begin{figure}[t!]
\center
	\includegraphics[width=0.55\textwidth]{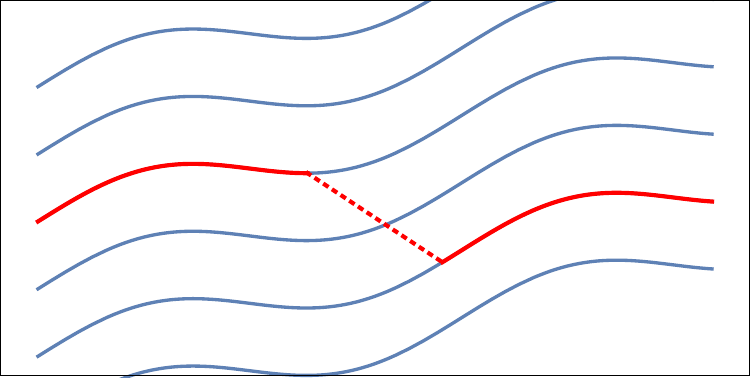}
	\caption{\it 
Various possible trajectories of the system evolving along $\phi$ at fixed values of $\chi^i$. Dynamical bifurcations during inflation correspond to transitions between different trajectories.}
	\label{fig:cartoon}
\end{figure}

 
For more general scalar potentials, the orthogonal directions will settle at (one or more) extrema of $V_{\rm eff}$ (see Fig.~\ref{fig:cartoon}).
The number and stability properties of these extrema can change during inflation, a phenomenon known  in dynamical systems as {\it bifurcations} \cite{strogatz:2000}. These bifurcations are elegantly captured by $V_{\rm eff}$. We will illustrate this using two characteristic examples from the recent literature. 

\subsection{Sidetracked inflation}	
Arguably the simplest setting that displays the bifurcation phenomenon is sidetracked inflation \cite{Garcia-Saenz:2018ifx}, originally formulated on a negatively curved space and a sum separable potential $V = U(\phi) +  {1 \over 2} m_h^2 \chi^2$, where $U$ is a single field potential corresponding to a variety of small-field inflationary models, including Starobinsky and natural inflation. The sidetracked phase succeeds the traditional slow-roll solution, after geometrical destabilization occurs, and so there is a transition from gradient flow slow-roll to a non-geodesic solution. Using a model with quadratic potentials and negative curvature:
 \begin{equation}
   \ud s^2 = \left(1 + {\chi^2 \over L^2} \right) \ud \phi^2 + \ud \chi^2\,,  \qquad     V = \tfrac12 m^2 \phi^2 + \tfrac12 M^2 \chi^2 \, ,
  \end{equation}
we will display the opposite phenomenon, i.e.~transition from non-geodesic to geodesic motion. 
As we will see, inflation takes place along $\phi$ and is thus perfectly suited to the effective potential framework. 

Let us first investigate the stability of the geodesic trajectory with $\chi = 0$. Particularly for quadratic potentials, both contributions to the isocurvature mass are approximately constant and read 
\begin{equation}
\mu^2 = M^2 - {2 m^2  \over 3 L^2 } \, .
\end{equation} 
Thus the curvature destabilizes the geodesic solution for 
 \begin{align} 
  L < \frac{\sqrt{2} m}{\sqrt{3} M} \,. \label{curvbound}
 \end{align}
However, for $\sqrt{3} M  L\lesssim \sqrt{2}m$, subleading corrections to the  isocurvature mass, consisting of the kinetic term for $\phi$ in the Hubble parameter,  become important and lead to bifurcations. In particular $\mu^2_s(\chi=0)<0$ at large $\phi$ and it slowly increases as inflation proceeds along the geodesic, becoming positive at
 \begin{align}
  \phi^2_{\rm cr} = \frac{4 m^2}{3(2m^2 - 3 L^2 M^2)} \,,
 \end{align}
where we have assumed $\phi >1$.  

\begin{figure}[t!]
 \center
	\includegraphics[width=1.1\textwidth]{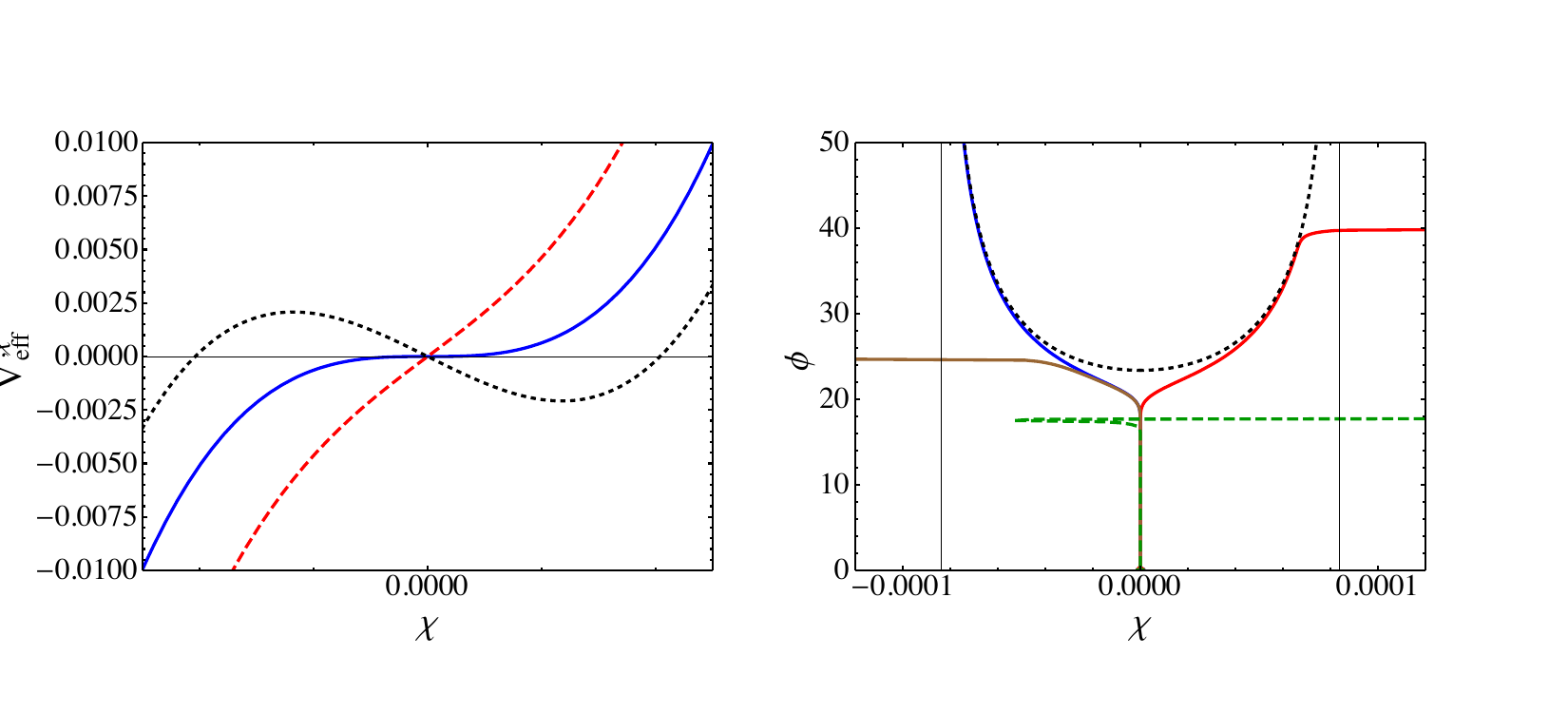}
	 \vspace{-0.8cm}
	\caption{\it 
{\rm Left:} The effective gradient of sidetracked inflation with $L=0.0034$, $m=1$ and $M=240$ along the stationary direction $\chi$ for different values of $\phi$, signaling the existence of one or three points of $V_{\rm eff}^{,\chi}=0$. The stability of each is determined by the slope of the curve. {\rm Right:} The corresponding bifurcation diagram. The black-dotted curve are the non-geodesic solutions  to Eq.~\eqref{stationary}, while the colored curves correspond to numerical solutions of the background system. }
	\label{fig:sidetracked}
\end{figure}

The subleading terms also determine the fate of the background trajectory when the geodesic solution is unstable. In addition to a local maximum, the subleading terms induce two minima in the effective potential at 
 \begin{align}
  \chi_{\pm}^2 =  L \left( \frac{\sqrt{2}m}{\sqrt{3}M} - L \right) \,, \label{nongeod}
 \end{align}
for $\phi \gg \phi_{\rm cr}$. The background trajectory will smoothly transit from the early non-geodesic trajectory at $\chi_{\pm}$ to the subsequent geodesic phase at $\chi =0$. Fig.~\ref{fig:sidetracked} shows the evolution of the effective gradient $V^{, \chi}_{\rm eff}$ and its zeroes as $\phi$ evolves, resulting in a pitchfork bifurcation. Moreover, it is clear from the figure that the numerical trajectories converge to the geodesic solution somewhat later; this can be understood as inertia in the moduli system, and indeed the different trajectories only become geodesic when $\mu^2_s \simeq H^2$ rather than $0$. 

\subsection{Hyperinflation}
A second example displaying a similar phenomenon is hyperinflation \eqref{disc}. It admits three solutions: radial evolution with $\theta' =0$, and two spiralling phases where the normalized angular velocity 
\begin{equation}
y = L \sinh \left({\rho \over L}\right) \theta' \, ,
\end{equation} 
is either zero or non-zero. For the first two $\epsilon \approx \epsilon_V$ and so $\omega \approx 0$. Stability of these solutions follows straightforward from our analysis. First, $M_{\rm eff}^2=0$ (as the effective gradient is identically zero $V^{,\theta}_{\rm eff}=0$), which is a consequence of the shift symmetry in $\theta$. Second, if we study the evolution of $y$ we obtain 
\begin{equation}
	y ' = - \left(3 - {1 \over 2} (\rho')^2 + {\rho' \over L} \coth\left({\rho \over L}\right)\right) y \, ,
\end{equation}
where the term in parenthesis is identical to Eq.~\eqref{eq:mvel}, after substituting $\epsilon \to \epsilon_V$ and $\rho' = -  (\log V)_{,\rho}$. When the gradient exceeds a critical value the gradient-flow solutions ($y=0$) become unstable and the system is driven towards the hyperinflation attractor.

For the simple example with
\begin{align}
   V = \tfrac12 m^2 \rho^2 \,,  \label{spherical}
  \end{align}
the trajectory undergoes  such a transition 
at $\rho = 2/(3L)$. Remarkably, one can bring all these solutions to proceed along a single direction via the field redefinition 
 \begin{equation} \label{eq:transformation}
 \cosh \left({\rho \over L}\right)  = \cosh \left({\chi \over L}\right) \cosh \left({\phi \over L}\right)  \, , \qquad  \cot(\theta) = \coth \left({\chi \over L}\right) \sinh \left({\phi \over L}\right) \, ,
 \end{equation}
leading to
 \begin{equation}
 \label{eq:hyperinsidetrackedcoordinates}
   \ud s^2 = \cosh^2 \left( {\chi \over L} \right) \ud  \phi^2 + \ud  \chi^2   \,.
 \end{equation}
This maps any spherically symmetric potential $V(\rho)$ onto a particular $V(\phi,\chi) $, providing all the necessary ingredients for realizing sidetracked inflation along $\phi$ \footnote{By ``sidetracked'' we refer to models that admit one geodesic solution along the minimum of the ``heavy'' field potential and two non-geodesic ones, generalizing the specific models of Ref.~\cite{Garcia-Saenz:2018ifx}.}.

\begin{figure}[t!]
	\center
	\includegraphics[width=1.01\textwidth]{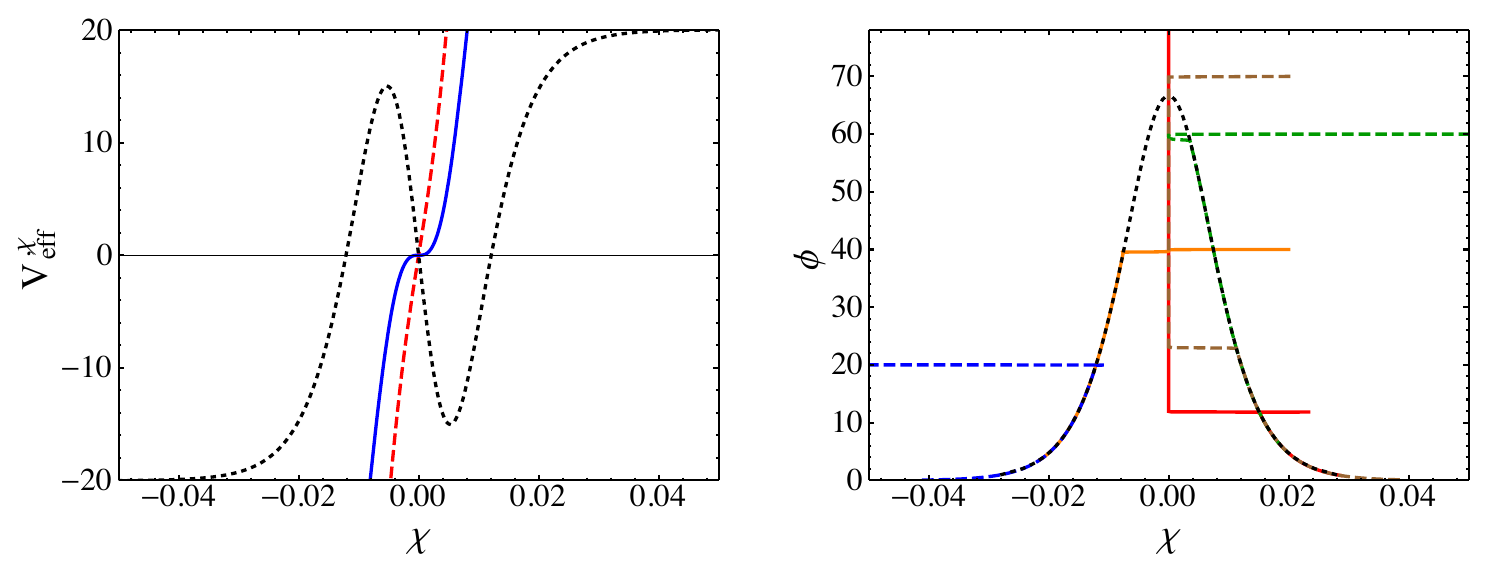}
	\vspace{-0.8cm}
	\caption{\it
		{\rm Left:} The effective  gradient  for hyperinflation in the coordinates of Eq.~\eqref{eq:hyperinsidetrackedcoordinates} at different $\phi$-values with $m=1$ and $L=0.01$.
		{\rm Right:} The corresponding bifurcation diagram. The black-dotted curves are the non-geodesic solutions to Eq.~\eqref{stationary}, while the colored curves correspond to numerical solutions of the background system.
	}
	\label{fig:hyper}
\end{figure}

Close to the geodesic solution  ($\chi=0$), the scalar potential reads (assuming $\phi > L$)
 \begin{equation}
  V = \tfrac12 m^2 \phi^2 + \tfrac12 m^2 {\phi \over L} \chi^2  \,.
 \end{equation}
The effective mass for motion along $\chi=0$ reads $M^2_{\rm eff} ={m^2\over L}(\phi -{2\over 3L} ) $,  becoming negative for $\phi< \phi_{\rm crit}=2/3L$.
At larger field values the geodesic solution is stable as the orthogonal field is strongly stabilised, while it becomes unstable at smaller field values. At this point, two new stable non-geodesic solutions come into existence, thus 
making up a pitchfork bifurcation \cite{strogatz:2000} (see Fig.~\ref{fig:hyper}).

\section{Generalizations and comparison} \label{sec:iso_pert}

\subsection{Beyond isometries}
 In Ref.~\cite{Bjorkmo:2019fls} a unification scheme was considered for the non-geodesic phase, based on the large turn rate of studied models. Stability of the solution  was shown using the perturbations' equations in the  adiabatic/entropic decomposition and requiring $\mu_s^2>0$,
where $\mu^2_s$ is the isocurvature effective mass
 \begin{equation}
  \mu^2_s = V^{,\chi}_{~~;\chi} + \epsilon R H^2 + 3 \omega^2 \,.
 \end{equation} 
However, as has been  pointed out in Ref.~\cite{Cicoli:2018ccr},  it is possible to have both a stable homogeneous solution and unstable orthogonal perturbations, leading to an apparent paradox. The resolution of this apparent paradox becomes clear if we compare $\mu_s$ with the stability criteria we presented earlier, in particular $M_{\rm eff}$
\begin{equation}
\mu_s^2 =  M_{\rm eff}^2 - \mathcal{G}^{\chi \chi} \Gamma^{\phi}_{\chi \chi} V_{,\phi}  + \epsilon H^2 R^{(\phi)} \, .
\end{equation}
Using $\chi'\approx 0$, $D=\left( \log \sqrt{{\cal G}}\right)'$ and
\begin{align}
D' &\approx {1 \over 2} {\mathcal{G}_{\chi\chi,\phi\phi} \over \mathcal{G}_{\chi\chi} } v^2 - {1 \over 2} \left( {\mathcal{G}_{\chi\chi,\phi} \over \mathcal{G}_{\chi\chi} } \right)^2v^2 \\
\epsilon R^{(\phi)} &= -{1 \over 2} {\mathcal{G}_{\chi \chi, \phi\phi} \over \mathcal{G}_{\chi\chi} }v^2 + {1 \over 4} \left({\mathcal{G}_{\chi \chi, \phi} \over \mathcal{G}_{\chi\chi} v}\right)^2 v^2 \, .
\end{align} 
we can rewrite the previous in a more geometrical way as\footnote{An interesting parallel exists between Eq.~\eqref{misoc} and Eq.~(5) of \cite{Achucarro:2018ngj} if one makes the substitution $D= -2h_i/H$. While both relations describe the mass of isocurvature modes, they were each derived in a different context. We do not fully understand their relation at this point, and thus leave this as an open question for future work.}:
\begin{equation} \label{misoc}
{\mu_s^2\over H^2}  \approx {M_{\rm eff}^2 \over H^2} - \left(3 -\epsilon + D \right) D - D' \, .
\end{equation} 
The two masses are equal when the metric has an isometry in the inflationary direction,
 which is the case for the examples in  \cite{Bjorkmo:2019fls}. Otherwise, $\mu_s^2$ and $M_{\rm eff}^2$ can differ and even have opposite signs. While this might sound surprising, the situation is similar to the familiar case of a spherically symmetric quadratic potential in flat target space. In polar coordinates $\mathcal{G}_{\chi \chi} = \phi^2$ and $V= \tfrac12 m^2 \phi^2$,  inducing a difference between both mass notions in \eqref{misoc}. The effective mass vanishes, indicating a range of neutrally stable trajectories on the attracting surface, while the isocurvature mass is positive, corresponding to a decrease of the proper distance between these trajectories, and a corresponding suppression of isocurvature fluctuations, as one approaches the minimum at $\phi = 0$.

The Hamilton-Jacobi formalism provides a clear illustration between the two (effective and isocurvature) mass notions in the absence of an isometry. The discussion around Eq.~\eqref{eq:HJ} holds for any metric of the form of Eq.~\eqref{general-metric} and thus generates an infinite set of adjacent, non-isolated critical points for the orthogonal fields. One can check that  $M^2_{\rm eff}=0$  for such constructions, highlighting the flat directions, while the isocurvature mass is proportional to the additional terms in \eqref{misoc}. For example, by choosing 
\begin{equation} 
  \ud s^2=\rho^2 \ud\theta^2+{\cal G}_{\rho\rho}(\theta)\ud\rho^2 \,, ~ V={m^2\over 2}\left(\theta^2 - {2\over 3\rho^2}\right) \,, 
  \label{eq:HJnonisometricpotential}
\end{equation}
the background trajectories of Ref.~\cite{Achucarro:2019pux} 
\begin{equation}
\rho =\rho_0\, , \qquad \dot\theta = \pm \sqrt{2/3} m/\rho_0^2\, ,
\label{eq:HJsol}
\end{equation} 
carry over, while the isometry along $\theta$ is broken. 
{One can check this by examining the background equations of motion
\begin{eqnarray}
\ddot \theta + 3H\dot\theta + {2\over \rho}\dot\rho \dot\theta - {1\over 2}{\partial_\theta {\cal G}_{\rho\rho}(\theta)\over \rho^2}\dot \rho^2 + {1\over \rho^2}V_{,\theta}=0
\\
\ddot \rho + 3H\dot\rho +{\partial_\theta {\cal G}_{\rho\rho}(\theta) \over {\cal G}_{\rho\rho}(\theta) }\dot\rho \dot\theta - {\rho\over {\cal G}_{\rho\rho}(\theta)}\dot \theta^2 + {1\over {\cal G}_{\rho\rho}(\theta)}V_{,\rho}=0
\end{eqnarray}
We can see that Eq.~\eqref{eq:HJsol} satisfies the equations of motion, since on the attractor $\dot\rho=0$ the term ${\cal G}_{\rho\rho}$ cancels out. The same holds for the slow-roll parameter and the Hubble parameter, while the turn rate is affected by the presence of ${\cal G}_{\rho\rho}$
\begin{equation}
\theta' = -{2 \over \theta \rho_0^2} \, , \qquad \epsilon = {2 \over \theta^2\rho_0^2}\, , \qquad H = {m \theta \over \sqrt{6}} 
\, , \qquad 
{\omega^2  }= {2\over 3}{ m^2\over \rho_0^2}  {\cal G}^{\rho\rho}
\, .
\end{equation}}

\begin{figure}[t!]
	\center
	\includegraphics[width=0.49\textwidth]{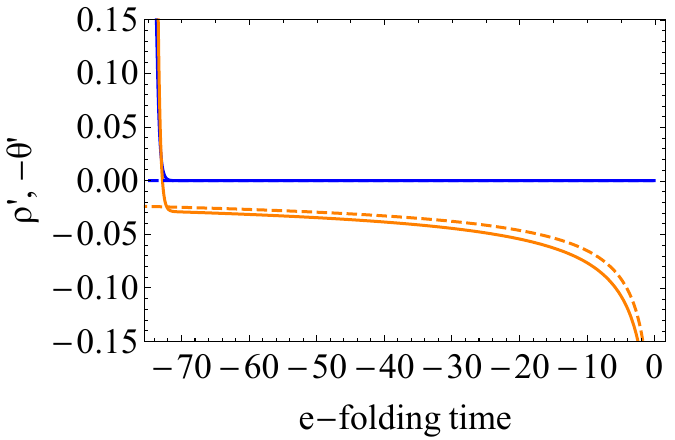}
	\includegraphics[width=0.49\textwidth]{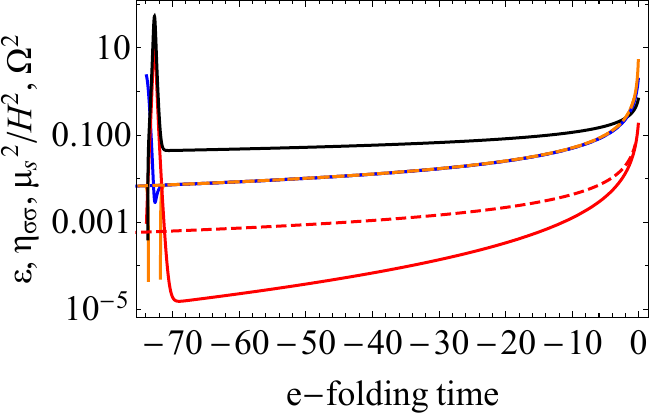}
	\caption{\it 
	Various dynamical quantities for the two models \eqref{eq:HJnonisometricpotential} with $\mathcal{G}_{\rho\rho}=e^{\theta}$ (solid lines) and $\mathcal{G}_{\rho\rho}=1$ (dashed lines) and initial conditions $\{\rho, \theta, \sqrt{\mathcal{G}_{\rho\rho}} \rho',  \sqrt{\mathcal{G}_{\theta\theta}} \theta' \}_{\rm init} = \{4,  4 , 1.5, 1.5\} $.
	 \newline
	Left: The velocities  $\rho'$ (blue) and $\theta'$ (orange).
\newline
	Right: The slow-roll parameters $\epsilon$ (blue), $\eta_{\sigma\sigma}$ (orange), turn rate $\Omega$ (red) and the isocurvature mass-squared $\mu_s^2$ (black). The effective mass for the second model reached the precision accuracy and was omited.
}
	\label{fig:HJ}
\end{figure}

While $M^2_{\rm eff}=0$, signaling the existence of background trajectories
for any constant value $\rho_0$, as long as Eq.~\eqref{eq:mvel} is satisfied,
the isocurvature mass $\mu_s^2$ can be either stabilizing or tachyonic. 
In the special case of $\mu_s^2=0$, isocurvature modes grow on super-horizon scales at a constant rate. Combined with a constant  turn rate, they continuously seed the adiabatic modes outside the horizon, leading to predictions that mimic those of  single-field models  \cite{Achucarro:2019pux}.
Let us we choose  a negatively curved manifold with
\begin{equation}
G_{\rho\rho} \sim e^{\theta/L} \, , \quad {\cal R} = -{1\over 2L^2 \rho^2} \, ,
\label{eq:GHJnonisometric}
\end{equation}
Even though the curvature of the manifolds given in Eq.~\eqref{eq:GHJnonisometric} is singular in the origin $\rho=0$, we can still view it as holding for $\rho>0$. For this model, the potential given in of Eq.~\eqref{eq:HJnonisometricpotential} is also singular at $\rho=0$, so the manifold must be smooth in the region of validity of the potential.
 The isocurvature modes in this model exhibit richer phenomenology compared to the flat metric case \cite{Achucarro:2019pux}, where $\mu_s=0$. In particular, along the (neutral) attractor at $\rho=\rho_0={\rm const.}$ the isocurvature mass is
 \beq
 {\mu_s^2 \over H^2} = {1 \over  L^2 \theta^2  \rho_0^4} \left ( 3 L\theta \rho_0^2 -1 \right ) \, .
 \eeq
 We see that $ \mu_s^2>0$ for $\theta > 1/L\rho_0^2$ and is negative otherwise. This means that the behavior of the isocurvature modes depends on the field-space curvature and the initial conditions, since different $\rho_0$ corresponds to different value of $\mu_s^2$. Furthermore, the character of the isocurvature modes can change during inflation, since $\theta$ is a monotonically decreasing function. The different behavior is shown in Fig.~\ref{fig:HJ}. Notice that for this model $\eta_{\sigma\sigma}= 6/(3 \theta^2 \rho^2 -3)$ and hence $\eta_{\sigma\sigma} \approx \epsilon$ for $\theta^2 \rho^2 \gg 1$. It is worth relating $\mu_s^2$ to Eq.~\eqref{eq:mvel}, which is a criterion for the existence of a stable solution with $\dot\rho=0$. This can be written as
\beq
 3 -\epsilon +{d\over dN}\log( \sqrt{{\cal G}} ) = 
 {1\over L \theta \rho_0^2} {\mu_s^2\over H^2} - {2\over \theta^2 \rho_0^2} >0
 \, ,
 \eeq 
hence no stable background trajectory exists for $\mu_s^2<0$.
This simple generalization of the model given in Ref.~\cite{Achucarro:2019pux} showcases the complex dynamics that can arise if one considers inflationary solutions with non-zero turn rate that do not proceed along field-space isometries.
For example a time varying value of $\mu_s^2$ and $\omega^2/H^2$, coming from a non-constant ${\cal G}_{\rho\rho}$ can lead to features in the scalar power spectrum.
 We leave a more thorough analysis of the phenomenology of such models for future work.

\subsection{Perturbations } 
Background trajectories with a non-zero turn rate can also affect the behaviour of fluctuations. We can always define gauge-invariant perturbations along the direction of motion ($Q_\sigma$) and perpendicular to it ($Q_s$).
On super-Hubble scales 
gauge invariant entropy perturbations $Q_s$ obey (see e.g. Ref.~\cite{Kaiser:2012ak} and references therein)
\begin{equation} \label{eq:eff}
	Q_s '' + (3 - \epsilon)Q_s' +  {\mu_s^2 \over H^2} Q_s \approx 0 \, .
\end{equation}
If $\mu_s^2>0$, $Q_s \rightarrow 0$,  allowing the comoving curvature perturbation $\mathcal{R}_c = Q_{\sigma}/\sqrt{2 \epsilon}$ to freeze at some point after horizon crossing. The moment of freeze-out  is mostly determined by the magnitude of $\mu_s^2/H^2$; this results from the slow-roll equations for $\mathcal{R}_c$ and $\mathcal{S} = Q_{s}/\sqrt{2 \epsilon}$ valid for $k\ll a H$ \cite{Kaiser:2012ak}: 
\begin{equation} 
\mathcal{R}_c ' \simeq  {2\omega \over H}  S \, , \qquad S ' \simeq \beta S \, ,
\end{equation}
where $\beta$ depends on slow-roll  quantities and $\mu^2_s/H^2$.

\begin{figure}[t!]
	\center
	\includegraphics[width=0.7\textwidth]{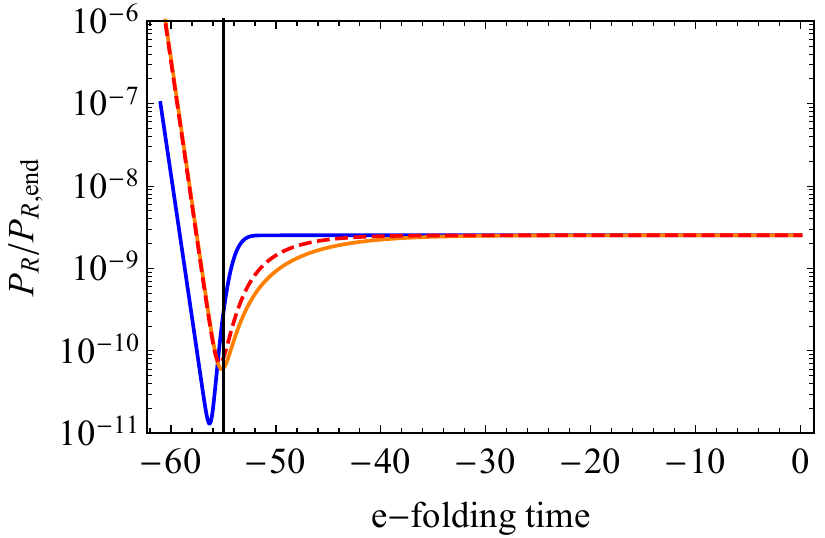}
	
	\caption{\it
		Power spectrum of curvature perturbation for angular inflation with $\alpha=1/600$ and mass ratio $(m_2/m_1)^2$ equal to 9 (blue), 100 (red) and 225 (orange).
	}
	\label{fig:ang}
\end{figure}

If after horizon crossing orthogonal fields are still evolving, then the non-uniqueness of the background trajectory is inherited by observables as well. One then finds a range of possible values for $\{n_s,r,...\}$ \footnote{Multi-field $\alpha$-attractors are exceptions to this rule, because the leading order contribution is independent of the specific initial state \cite{Achucarro:2017ing,Christodoulidis:2018qdw}.}. 
Analytical estimates can be constructed in the following way: on sub-Hubble scales one can identify uncoupled perturbations by an appropriate time-dependent rotation \cite{Cremonini:2010ua}; close and prior to horizon crossing, if the mass of isocurvature perturbations on sub-Hubble scales
\begin{equation}
m_s^2 \equiv \mu_s^2 - 4 \omega^2 \, ,
\end{equation} 
is large enough then $\mathcal{S}$ will be stabilized at a zero of its `effective gradient' given by \cite{Tolley:2009fg,Achucarro:2012sm}
\begin{equation}
\left({k^2 \over a^2} + m_s^2 \right) \mathcal{S} + 2 \omega \dot{\mathcal{R}}_c = 0\, .
\end{equation} 
Substituting this solution into the equation for $\mathcal{R}_c$ provides an equation similar to single-field but with a k-dependent sound speed. Note that in general solving these equations is a model-dependent problem. For example, in angular inflation with  $\alpha\ll1$ when the ratio of the heavy to light field is  $(m_2/m_1)^2 \lesssim {\cal O}(10)$, the curvature perturbation freezes shortly after horizon-crossing; when the masses of the two fields differ significantly then $|\beta| \ll 1$ and there can be substantial super-horizon evolution \cite{ang2}. Note that in both cases the background trajectory is unique, given by minimizing $V_{\rm eff}$, but perturbations behave differently, as shown in Fig.~\ref{fig:ang}.

\section{Summary and discussion} \label{sec:conclusion}

Multi-field models often display a strong attracting behaviour; orthogonal fields are stabilised by their effective potential, consisting of potential energy and generalized centrifugal forces due to non-geodesic motion. This can be interpreted as the (partial) minimisation of the total energy density given by the Hubble parameter as a function of the orthogonal fields and is the analogue of moduli stabilization, albeit on a time-dependent background. Moreover, as inflation proceeds, the stabilisation pattern can undergo pitchfork bifurcations, with a stable minimum becoming unstable with the simultaneous appearance of two new stable trajectories or vice versa. 
The total number of stable minus unstable solutions remains constant, characteristic of  pitchfork bifurcations. This structure is reminiscent of the waterfall transition  in hybrid inflation \cite{Linde:1993cn}.  It would be interesting to investigate whether the analogy to hybrid inflation can be extended beyond the background evolution, providing distinct observational signatures \cite{Clesse:2015wea, Lyth:2012yp, Guth:2012we} for  multi-field models exhibiting pitchfork bifurcations during inflation.

We presented a unifying perspective on different scenarios of multi-field inflation in curved geometries, based on the dynamical properties of the inflationary evolution after the decay of the initial transient regime. 
While angular inflation has a unique minimum of $V_{\rm eff}$, 
 both sidetracked and hyperinflation 
 exhibit  dynamical  pitchfork bifurcations. 
 This instability is therefore intrinsically of the same nature; analyzing hyperinflation after the coordinate transformation of Eq.~\eqref{eq:transformation} makes it a special case of sidetracked inflation. This  connects two models that were so far thought to be distinct, thus underlining the unifying nature of our approach. Moreover, it demonstrates that the conservation of angular moment is not essential to the bifurcation in hyperinflation.

Finally, we showed that the existence of an isometry along the inflationary direction is not a necessary condition for the existence and stability of inflationary attractors with a non-zero turn-rate. By providing a simple generalization of shift-symmetric orbital inflation \cite{Achucarro:2019pux}, we constructed a model in which the effective mass $M_{\rm eff}$ is identically zero for all members of a continuous family of trajectories with a constant radius, thus extending the notion of a neutrally stable attractor. However, the isocurvature mass on any of these trajectories is not zero but rather positive and depends on the field-space curvature. Furthermore $\mu_s^2$ evolves in time, allowing for the generation of features in the scalar power spectrum. We leave an extensive analysis of the intruiging phenomenology of inflationary models with broken isometries for future work.

\section*{Acknowledgements} The authors gratefully acknowledge stimulating discussions with Ana Ach\'ucarro, Cliff Burgess, Michele Cicoli, {Veronica Guidetti}, Sonia Paban, Gonzalo Palma, Robbie Rosati, {Francisco Pedro} and Vincent Vennin as well as financial support from the Dutch Organisation for Scientific Research (NWO).

\appendix

\section{Slow-roll conditions and stability}

\subsection{Adiabatic slow-roll conditions} \label{app1}
The potential slow-roll conditions for one field do not generalize in a straightforward manner, i.e.~simple conditions involving only (covariant) derivatives of the potential, in multi-field inflation. However, we can still derive some useful consistency relations involving kinematic quantities without solving for the equations of motion. To derive these relations we work in the standard adiabatic/entropic decomposition \cite{Kaiser:2012ak,  Peterson:2011yt, Renaux-Petel:2015mga, Achucarro:2012sm, Gong:2011uw}. Along the direction of motion the adiabatic field 
\beq
\dot{\sigma}^2 \equiv \mathcal{G}_{IJ}\dot{\phi}^I\dot{\phi}^J
\eeq
satisfies the following equation of motion
\begin{equation}
\ddot{\sigma} + 3 H \dot{\sigma} + V_{\sigma} =0 \, ,
\end{equation}
where the unit vectors $\hat \sigma^I$ along the adiabatic direction are defined as 
\begin{equation}
 \hat{\sigma}^I \equiv { \dot{\phi}^I \over \dot{\sigma}} 
 \end{equation}
and the corresponding potential gradient is
\beq
   V_{\sigma} \equiv V_{,I} \hat{\sigma}^I  \, .
\eeq
The first slow-roll parameter can be written as
\begin{equation}
\epsilon \equiv -{\dot H\over H^2} =  { \dot{\sigma}^2 \over 2 H^2 } \, .
\end{equation}
Assuming that $\epsilon \ll 1$ then imposing the condition $|\eta| \ll 1$ is equivalent to the smallness of 
\begin{equation}
\eta = 2\epsilon + 2{{\ddot{\sigma} \over H \dot{\sigma}}} \, .
\end{equation}
This implies that the $\sigma$ field follows its potential gradient, satisfying the slow-roll equation $3 H \dot{\sigma} + V_{\sigma}  \approx 0$. Finally, with $\dot{\sigma} \approx \dot{\sigma}_{SR}$ we obtain
\begin{equation}
\eta \approx 4 \epsilon - 2 \eta_{\sigma \sigma} \, ,
\end{equation}
where we defined the second  slow-roll parameter
in the adiabatic direction as
\begin{equation}
\eta_{\sigma \sigma} \equiv { V_{\sigma \sigma} - \omega^2 \over 3 H^2} \, .
\label{eq:etasigmasigmadef}
\end{equation}
It was first pointed out in Ref.~\cite{Christodoulidis:2018qdw} and further elaborated  in Ref.~\cite{Chakraborty:2019dfh} that slow-roll inflation in the mutli-field case requires $\eta_{\sigma\sigma}$ as defined in Eq.~\eqref{eq:etasigmasigmadef} to be small. In the single field case, this reduces to the usual form $\eta_{\sigma\sigma}\to V_{\sigma\sigma}/V$, which is not necessarily small in the presence of a large turn-rate.

In our coordinate system $\dot{\chi}=0$ and hence the adiabatic and orthogonal directions are
\begin{equation}
\hat \sigma^I = \left( \sgn \left( \dot{\phi} \right) \sqrt{G^{\phi\phi}}, 0 \right) \, , \qquad \hat s^I = \left( 0, -\sgn \left( \dot{\phi} \right) \sqrt{G^{\chi\chi}}, 0 \right) \, .
\end{equation}
Calculating $\epsilon$ and $\eta_{\sigma \sigma}$ for our coordinate choice and imposing slow-roll conditions on $\phi$ leads to Eq.~\eqref{eq:sr}.

\subsection{Stability method for background motion} \label{app}

The first step to investigate stability for a system of second order differential equations is to experss it in first order form with the definition of velocities as new variables $v^I \equiv (\phi^I)'$. Next, we notice that the condition for prolonged inflation $\epsilon' \ll 1$ suggests that the appropriate set of variables is $\{\phi,\chi, y, x \}$, where $y=\sqrt{\mathcal{G}_{\phi\phi}}\phi'$ and $x=\sqrt{\mathcal{G}_{\chi\chi}}\chi'$ are the normalized velocities. These are finite quantities and are almost constant during inflation. For a generic two-field metric 
\begin{equation}
\ud s^2 =  \mathcal{G}_{\phi \phi}(\phi,\chi) \ud \phi^2 + \mathcal{G}_{\chi \chi} (\phi,\chi) \ud \chi^2 \, ,
\end{equation} 
the dynamical system becomes
\begin{subequations}
	\label{dx4Dniso}
	\begin{eqnarray}
\phi' & = & { y \over \sqrt{\mathcal{G}_{\phi\phi}} } \, , \\
\chi' & = & { x \over \sqrt{\mathcal{G}_{\chi\chi}} } \, ,\\
	y' &=& - (3-\epsilon)  \left( y + {p_{\phi} \over \sqrt{ \mathcal{G}_{\phi\phi}} } \right) - { \mathcal{G}_{\phi\phi ,\chi} \over 2   \mathcal{G}_{\phi\phi} \sqrt{ \mathcal{G}_{\chi\chi}} }x y + {  \mathcal{G}_{\chi\chi ,\phi} \over 2 \mathcal{G}_{\chi\chi }  \sqrt{ \mathcal{G}_{\phi\phi}}  }  x^2 \, ,
\\
	x' &=& - (3-\epsilon)  \left(x + {p_{\chi} \over \sqrt{ \mathcal{G}_{\chi\chi}} } \right) - { \mathcal{G}_{\chi\chi ,\phi} \over 2   \mathcal{G}_{\chi\chi} \sqrt{ \mathcal{G}_{\phi\phi}}} x y + {  \mathcal{G}_{\phi\phi ,\chi} \over 2 \mathcal{G}_{\phi\phi }  \sqrt{ \mathcal{G}_{\chi \chi}}  }  y^2 \, . 	\end{eqnarray}
\end{subequations}
where $p_{I} = (\log V)_{,I}$. Linearization of the previous system around a stable background solution provides the stability criteria. In order to obtain some closed-form expressions we need to be able to ommit certain subdominant terms. Assuming the metric \eqref{general-metric} the inflaton's velocity is found constant
\begin{equation}
\uD_N v \approx 0 \Rightarrow v' \approx 0 \, ,
\end{equation}
and the equations of motion imply that the gradient (with upper index) of the inflaton field should be constant as well
\begin{equation}
 v \approx - p^{\phi} = {p_{\phi} \over \mathcal{G}_{\phi \phi}}  \, .
\end{equation}
Since by assumption the $\chi$ field is frozen, we arrive at the requirement $p_{\phi}'\approx 0$ or $p_{\phi,\phi} \approx 0$. Therefore, the stability conditions given in Section~\ref{sec:stab} will be similar to those derived in Ref.~\cite{Christodoulidis:2019jsx}, up to slow-roll suppressed corrections.

\end{document}